\documentclass[manuscript,nonacm]{acmart}

\usepackage{algorithm}
\usepackage{algpseudocode}
\usepackage{wrapfig}
\AtBeginDocument{%
  \providecommand\BibTeX{{%
    \normalfont B\kern-0.5em{\scshape i\kern-0.25em b}\kern-0.8em\TeX}}}

\begin{document}

\title{BrickPal: Augmented Reality-based Assembly Instructions for Brick Models}

\author{Yao shi}
\author{Xiaofeng Zhang}
\author{Ran zhang}
\author{Zhou Yang}
\author{Xiao Tang}
\author{Hongni Ye}
\author{Yi Wu}
\begin{abstract}

The assembly instruction is a mandatory component of Lego®-like brick sets.
The conventional production of assembly instructions requires a considerable amount of manual fine-tuning, which is intractable for casual users and customized brick sets.
Moreover, the traditional paper-based instructions lack expressiveness and interactivity.
%
To tackle the two problems above, we present BrickPal, an augmented reality-based system, which visualizes assembly instructions in an augmented reality head-mounted display. It utilizes Natural Language Processing (NLP) techniques to generate plausible assembly sequences, and provide real-time guidance in the AR headset.
%
Our user study demonstrates BrickPal's effectiveness at assisting users in brick assembly compared to traditional assembly methods. Additionally, the NLP algorithm-generated assembly sequences achieve the same usability with manually adapted sequences.
\end{abstract}

 \begin{teaserfigure}
    \centering
   \includegraphics[width=0.75\textwidth]{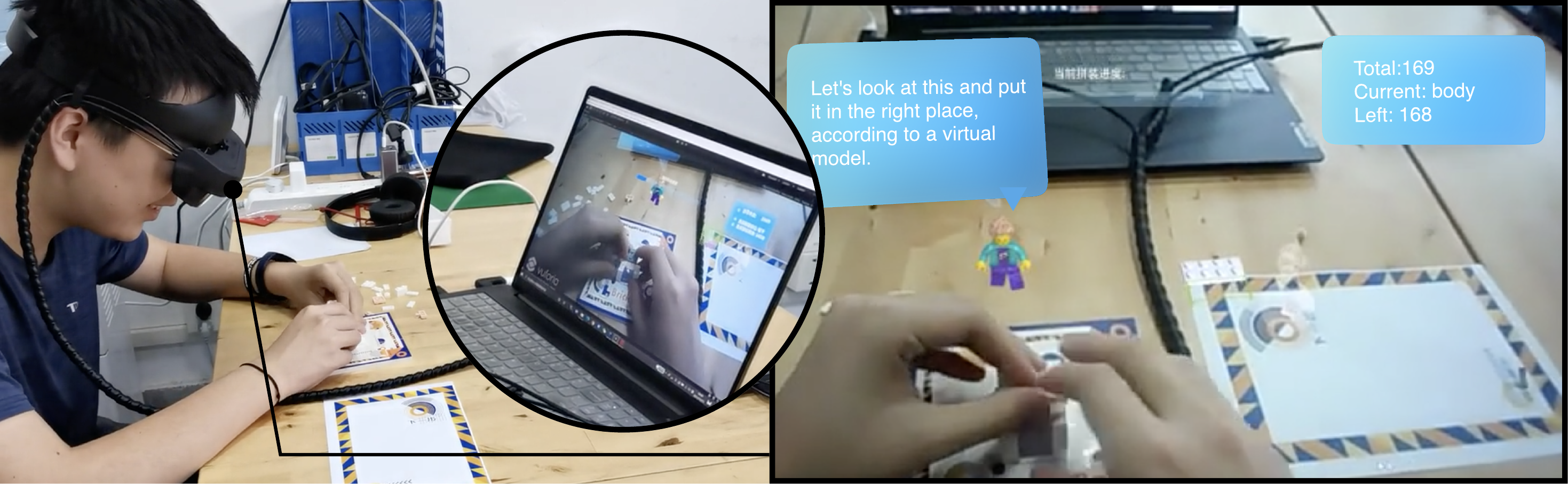}
   \caption{Assemble bricks with BrickPal. A user wears an AR Head-mounted Device (HMD) to assemble bricks by following Natural Language Processing-generated instructions.}
   \Description{.}
   \label{fig:teaser}
 \end{teaserfigure}

\maketitle
\section{Introduction}
An assembly indicates joining multiple parts to a target form or functionality.
The assembly process dramatically enhances the flexibility of product manufacturing.
It is achieved by splitting the entire fabrication into multiple sub-fabrication processes for different smaller-scale parts, which allows manufacturing the products in a faster, more asynchronous way.
Hence, assembly is a typical process in manufacturing various objects, including architectural structures, industrial machines, and consumer goods.
For most industrial use cases, assemblies happen in factories.
However, for consumer goods, such as IKEA® furniture and Lego® brick sets, the end-users conduct the actual assembly tasks.
Suppliers usually provide paper-based manuals with pre-generated instructions to help the user correctly assemble sophisticated objects.
However, the pre-generated instructions are designed for single deterministic assembly sequences.
The generation of these sequences usually requires the amount of effort, and lacks flexibility for open-ended assembly, such as toys involving multiple sequences.
Additionally, paper-based media restricts the instructions to 2D illustrations,
which could be hard to infer for complex 3D shapes.

In this paper, we propose a solution that tackles both of the problems above for Lego®-like brick models.
In our approach, we visualize the assembly instructions by displaying an overlay using a commercially available Augmented Reality (AR) headset.
Using real-time marker-based tracking, users are able to see the 3D-projected assembly guidance in-situ.
Alongside the augmented reality interface, we utilize machine learning techniques initially used in Natural Language Processing (NLP) to generate the assembly sequences for any input brick models automatically.
Based on these techniques,
we presented BrickPal, an AR-based Smart Assembly guiding system for brick models, see Fig. \ref{fig:teaser}. In the following sections, we motivat our system design from existing related work (Section \ref{sec:related_Work}). We explain the implementation of the BrickPal prototype (Section \ref{system_design}). Finally, we demonstrate the effectiveness and the enhanced experience by numerical evaluation (Section \ref{sec:nlp_eval}) and a user study comparing our system with paper-based instructions and manually generated assembly sequences (Section \ref{sec:user_study}).

\section{Related Work}
\label{sec:related_Work}
\subsection{AR Instructions for assembly tasks}
Augmented Reality (AR) is a term of technology that allows people to view the real-world environment as an “augmented” form by overlaying computer-generated virtual elements or objects~\cite{Krevelen2010ASO}. 
\citet{Bottani2019AugmentedRT} identified the main areas and sectors where AR is currently deployed. Their results confirmed that assembly and maintenance were the most popular application fields of AR during past decades.
\citet{EggerLampl2019AssemblyIW} studied the effect of AR on the work experience in industry 4.0 contexts. Their results showed that interactive AR assembly instructions positively impacted perceived utility and satisfaction.
As a more specific example, ProcessAR~\cite{chidambaram2021processar} successfully vitalized real-world tools and assisted users to assemble simple objects without transitioning between different modalities and interfaces.
For assembling brick models, \citet{DBLP:journals/corr/abs-1907-12549} presented a marker-based AR assembly system for Lego® architectures.
This system overlays the next brick on top of the actual brick model and provides real-time visualization in the AR headset.
We follow the same idea of visualizing assembly sequences using an AR headset. However, instead of showing a manually generated assembly sequence, we use machine learning techniques to provide just-in-time assembly guidance.

%
\subsection{Transformer-based generative tasks}
Recently, large-scale pre-trained language models (PLMs), e.g., BERT~\cite{DBLP:conf/naacl/DevlinCLT19} and GPT~\cite{radford2019language}, have achieved great success in NLP. 
The core structure of PLMs is the Transformer module~\cite{DBLP:conf/nips/VaswaniSPUJGKP17}, which can capture long-range interactions between different words.
The long-range memory capability of the Transformer module makes it possible to deal with assembly and composition tasks.
Our main observation is that the assembly sequence is an ordered list of discrete tokens. The order of the tokens is highly related to the interactions with other tokens in the list.
This is analogous to NLP problems as sentences are ordered word lists, in which words interact with each other.
Along this direction, researchers have proposed various preliminary results. \citet{DBLP:conf/cvpr/RibeiroBCP20} utilized the Transformer architecture for encoding sketches and showed its effectiveness for downstream classification and reconstruction tasks. 
%
\citet{DBLP:conf/nips/CarlierDAT20} took the full Transformer architecture to generate Scalable Vector Graphics (SVG) but in a non-autoregressive fashion. 
\citet{DBLP:conf/icml/NashGEB20} proposed PolyGen, a Transformer decoder autoregressive model, to generate 3D meshes. 
\citet{DBLP:journals/corr/abs-2105-02769} extended PolyGen in order to generate CAD (Computer-Aided Design) sketches.
%
These attempts utilized Transformer to generate suitable constructive data and provide brilliant results.
However, in our algorithm, we do not only adopted Transformer to model the token order sequences, but also generate positional information. 
This is a special research approach since the traditional Transformer model only supports discrete tokens but not continuous positional information.
We will explain how we achieve this target in the following sections.
%
\section{The BrickPal System}
\label{system_design}
\begin{figure}[!tbp]
    \centering 
    \includegraphics[width=0.85\linewidth]{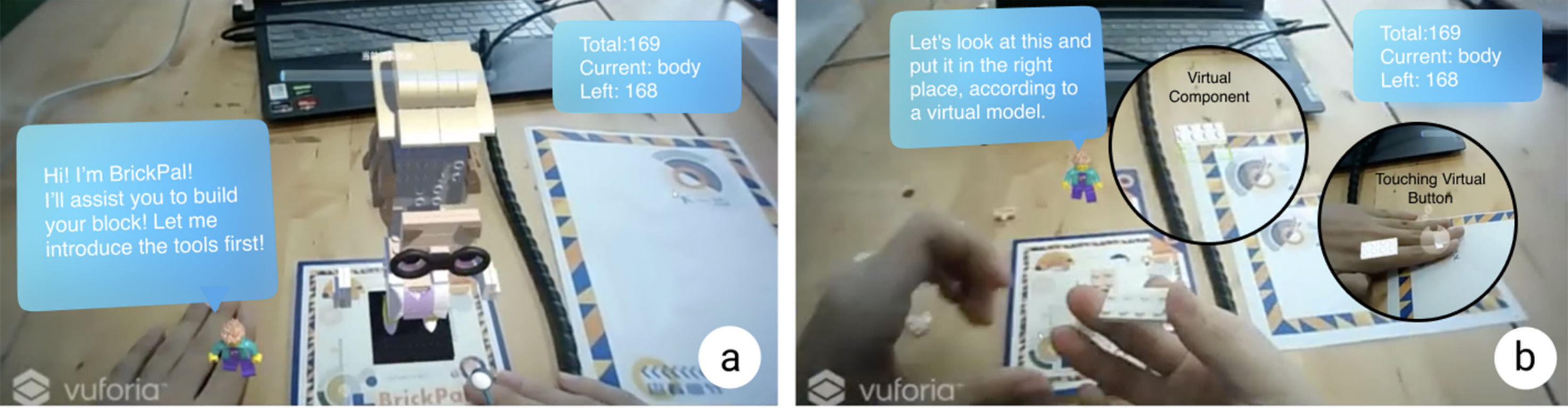}
    \caption{Walkthrough of BrickPal.(a) Initial step, the virtual assistant introduces the system; (b) Assembly, a virtual brick overlay indicates the next step, the user finds the target brick and assembles it and continues by touching a virtual button.}
    \label{fig:prototype}
    \vspace{-1.5em}
\end{figure}
\subsection{Prototype}
We designed BrickPal, a system that exploits marker-based AR and NLP algorithms to enhance the functional and emotional properties of brick model assembly guidance.
The system starts with an introduction by a virtual assistant. As shown in Fig.~\ref{fig:prototype}a, when the user puts on the AR HMD device,
the image-based marker detection runs automatically to track the pedestal of the brick model.
Then a virtual assistant introduces a preview of the final model and guides the user to try the magnifier tool (a virtual button with a magnifier metaphor). 
In the assembly session, a rendered model of the potential brick is floating in the air. It guides the user to search for the corresponding physical one. The target position indicates the placement as well, see Fig.~\ref{fig:prototype}b. 
Once the user has finished the current step, one can update the next brick hint by touching the ``Next Step'' marker.
Contrary to a traditional manual, the user does not need to stop to search for the next step by referring to the handbook.
More importantly, the step-by-step guidance in our system is generated by a machine learning-based sequence ordering algorithm, see Section~\ref{sec:sequence}.
Benefiting from the fast machine learning-based sequence ordering algorithm, our prototype supports just-in-time generation of assembly guidance.
Thus, we provide versatile routine modes for the user interaction, including: 
1) Single-track assembly sequence computed from the brick model,
2) Multi-track assembly guidance that allows the user to pick the brick from a few potential candidates and get the corresponding guidance in real-time, 3) A pair-assembly mode, where the user can enable the guidance on demand. 
In the following sections, we will explain the implementation of our prototype.


\subsection{Transformer-based Sequence Ordering Algorithm}
\label{sec:sequence}
\subsubsection{Problem definition}
\label{sec:def}
In this part, we introduce how NLP helps to generate flexible assembly sequences.
The problem studied in this paper is the assembly sequence ordering for brick models. Given a brick model, the bricks in this model can be noted as $T = \{t_1, t_2, t_3, ..., t_n\}$ where $n$ represents the total number of the bricks. For each brick, their spatial position (a 3-dimensional vector) can be noted as $p_i$ where $i$ represents the index of the brick. The objective of our problem is to give an ordered sequence $\{t_{(1)},t_{(2)},...,t_{(n)},\}$. Following the ordered sequence, the brick model can be assembled without brick overlapping, collision, or overhanging.

\subsubsection{Methodology}
We propose a 3D positional-augmented language model named 3D-PALM. Our proposed method can integrate the 3-dimensional information into a masked language model (MLM). This allows the language model to learn the brick sequence distribution in considering the spatial information distribution for not creating collisions and overlaps.


The targeted problem is analogous to text generation tasks in NLP. The probability distribution of the next generated brick is calculated based on the existing brick sequence. It is thus feasible to give several probable positions to assemble at each prediction step. The user can pick one brick out of the possible candidates. This property makes the assembly more flexible and enjoyable. 

\begin{figure}[!tbp]
    \centering
    \includegraphics[width=0.8\linewidth]{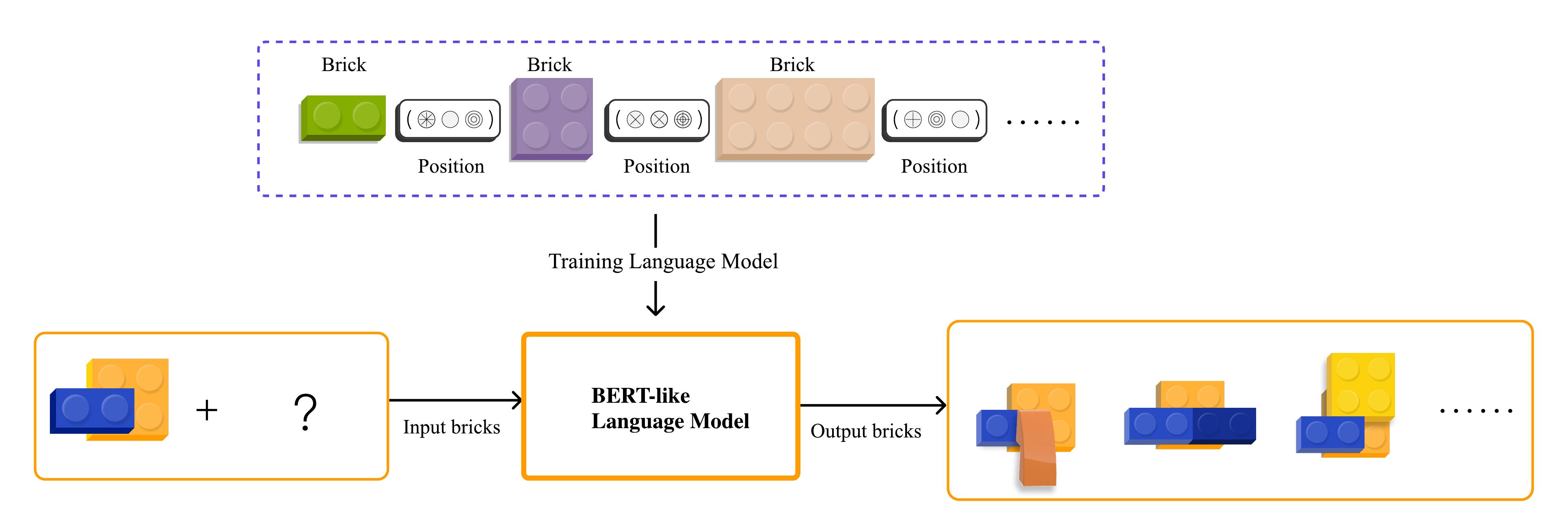}
    \caption{Overview of 3D-PALM. At the top of the figure, the BERT-like Language Model is trained by the known brick+position sequences. At the bottom of the figure, with the trained LM, we can give the next possible bricks' distribution. Several possibilities can be compatible with the brick model.}
    \label{fig:input_seq_change}
    \vspace{-1.5em}
\end{figure}

Despite the analogy, the assembly sequence ordering problem has one significant difference. The brick's position is 3-dimensional rather than 1-dimensional for a word in a sentence. 
The 1-dimensional position embedding of the original MLM cannot adequately describe the 3-dimensional spatial position of each brick. It may cause spatial conflicts among different bricks in the ordered assembly sequence.
The input of the MLM is originally a sequence consisting of bricks only. 
As shown in Fig. \ref{fig:input_seq_change}, we insert a position token representing the relative position between each two adjacent bricks in the brick sequence. Thus, the input sequence changes to $\{t_1, pr_1, t_2, pr_2,...,t_i, pr_i,...,t_{n-1},pr_{n-1},t_n\}$ where $t$ is the original brick input sequence and $pr$ is the relative position calculated by $pr_i = p_{i+1} - p_i$.

We then adopt a discretization process.
For each dimension of a position vector, the bricks possess an intrinsic length $[l_1,l_2,l_3]$, i.e., the standard size of bricks' three sides. The relative position can be discretized by dividing each dimension's value by the corresponding intrinsic length. This process can be formally defined as:
\begin{equation}
\label{eq:discretization}
    pd_{i,j} = \max( \min(\lfloor d_{i,j}/l_j + l_{\rm{max}}/2 \rfloor, l_{\rm{max}}), 0), i \in [\![1,n]\!], j \in [\![1,3]\!] \ \text{.}
\end{equation}
$l_{\rm{max}}$ denotes the max discretization number in the discretization process, $\lfloor \cdot \rfloor$ denotes the floor function, $i$ denotes the index of bricks, $j$ denotes the index of dimensions. $l_{\rm{max}}$ could be settled with different values to control the size of vocabulary considered. With Equation~\ref{eq:discretization}, we can consider both the positive and the negative values of $pr_i$.

For each dimension, the distance is split into $l_{\rm{max}}$ categories. In total, the relative position can be transformed into $l_{\rm{max}}^3$ discrete classes. For each class, a special position token is attributed. 

\subsection{Implementation}

\label{sec:S_A_I}

We developed the BrickPal prototype with Unity~\cite{unity} engine.
More specifically, we use the assets from Unity Lego® Microgame~\cite{lego-microgame} template for the model of virtual assistant and its animations.
For the tracking and interaction with brick models in the AR HMD, we apply the image-based marker tracking and virtual button feature from Vuforia SDK~\cite{vuforia}.
We run our 3D-PALM algorithm on a Flask~\cite{flask} server, which retrieves the updated assembly sequence from the Unity Engine and sends back possible candidates through HTTP requests.

For the sequence ordering algorithm, we implement BERT-like language model using Huggingface Transformers Library~\cite{wolf-etal-2020-transformers} (the code will be available upon publication).
In total, the 3D-PALM consists of a 6-layer Transformer encoder. For each self-attention layer, 12 heads are used for contextual tokens. The batch size is set to 32. We trained our language model on a dataset with the assembly sequences generated by a carefully implemented rule-based searching method for commercial brick sets. 
\section{Technical Evaluation}
\label{sec:nlp_eval}
To demonstrate the effectiveness of the BrickPal system and its positive impact on users' experience, we designed a two-fold evaluation for the prototype.
The first part is a technical validation of the NLP-based assembly sequence generation algorithm. We evaluate the algorithm with multiple quantitative metrics for machine learning models.
Accompanying with the technical evaluation, in Section~\ref{sec:user_study}, we conducted a group user study,
and collected the quantitative and qualitative feedback from the participants.


\subsection{Experimental settings}
We utilized 132 commercial brick models to test our BrickPal system. For each brick model, an ordered sequence is supplied by a rule-based algorithm with pre-defined constraints.
%
To validate the training of the 3D-PALM model, we monitored the perplexity (the lower the better) using different learning rate. The perplexity metric indicates the confidence of predictions.
%
Furthermore, we utilized widely used BLEU and ROUGE scores (the higher the better) to evaluate the generation quality.The BLEU and ROUGE approximately represent respectively Precision and Recall for the task~\cite{lin-2004-rouge,callison-burch-etal-2006-evaluating} .


\vspace{-1em}
\begin{wrapfigure}{r}{0.4\textwidth}
\vspace{-2em}
    \centering
    \includegraphics[width=\linewidth]{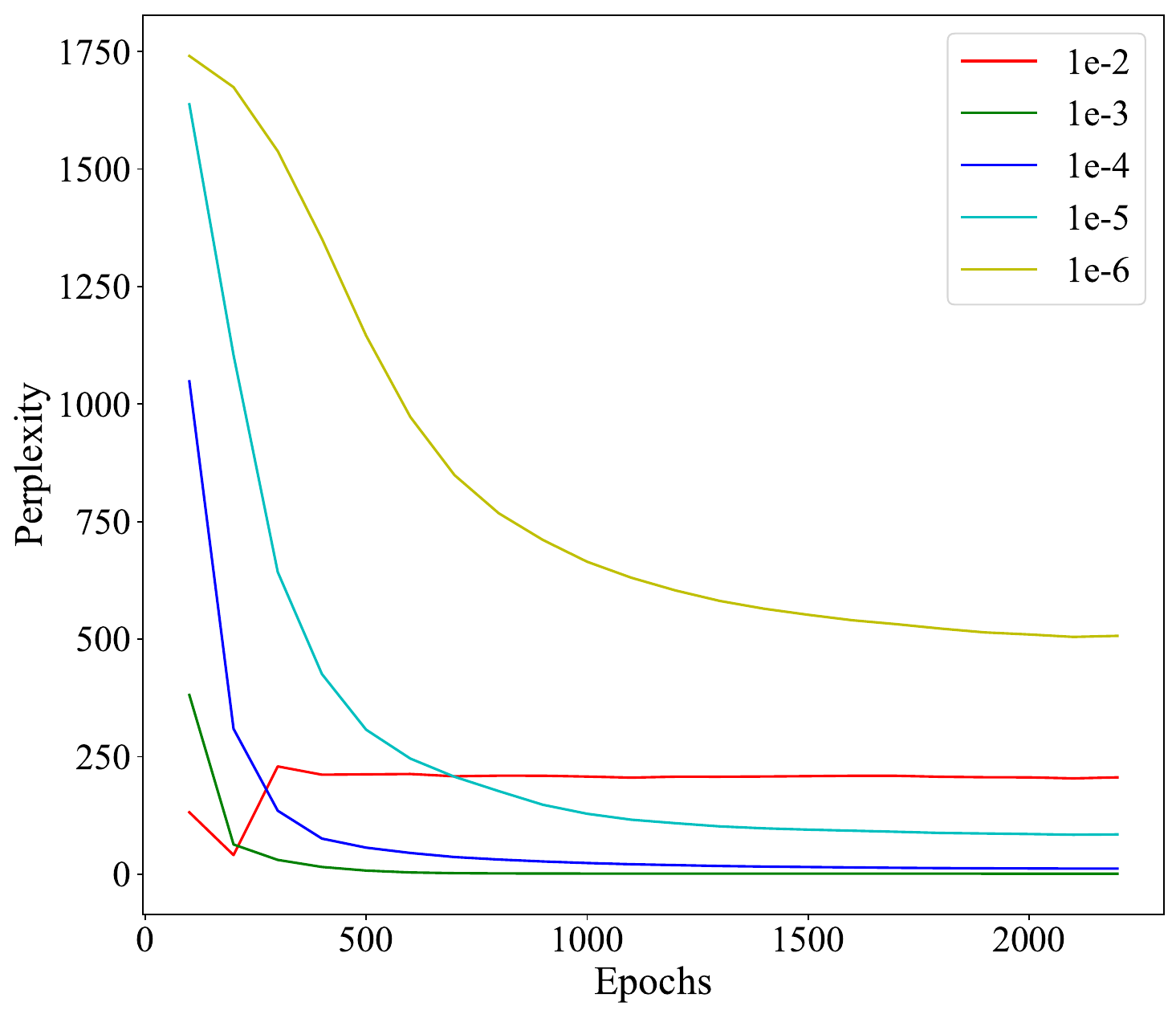}
    \vspace{-2em}
    \caption{Perplexity of the model w.r.t learning rates.}
    \label{fig:lr_ppl_epoch}
    \vspace{-2em}
\end{wrapfigure} 

\subsection{Experimental results}
As shown in Fig. \ref{fig:lr_ppl_epoch}, we observed that the best perplexity of 1.012 is attained with learning rate 1e-3. 
It means that whenever the model is trying to guess the next word it is as confused as if it had to pick between 1.012 words. This indicates an approximate certainty of the prediction is relatively good. 

To evaluate the generation performance, we defined a conditional generation task. Given a certain number of pre-assembled bricks (from 10 to 40), we use the trained masked language model to predict the next 80 bricks. And compare the result with the rule-based generation. The result is shown in Table~\ref{tab:gen_known}.

We also evaluated the generation time of our proposed 3D-PALM. We implemented a rule-based searching algorithm for the assembly sequence generation task, it needs 5 to 10 minutes to generate the full sequence for a brick model set with a hundred pieces, while our model consumes only 50 to 70 seconds (result in real-time updates for each step). Although the generation time is drastically reduced, the generation quality is still satisfactory, see Section \ref{sec:user_study}.


\begin{table}[!htbp]
    \centering
    \caption{BLEU and ROUGE scores on the conditional generation task, the higher the better}
    \label{tab:gen_known}
    \begin{tabular}{ccccc} \toprule
        Pre-assembled bricks & BLEU-4(\%) & ROUGE-1(\%) & ROUGE-2(\%) & ROUGE-L(\%) \\ \midrule
        10 & 0 & 33.91 & 16.19 & 33.28 \\  
        20 & 10.22 & 39.21 & 21.51 & 38.26 \\
        30 & 20.43 & 30.52 & 11.55 & 28.41\\
        40 & 15.10 & 25.36 & 8.27 & 24.22 \\ \bottomrule
    \end{tabular}
\end{table}



\section{User study}
\label{sec:user_study}
We experimented with investigating the impact of BrickPal system on the brick assembly tasks. Using the same set of the brick model with 22 bricks to assemble
, we developed three types of instructions, i.e., an AR guidance tool with flexible sequences, the same apparatus with manually created deterministic sequences, and a paper-based instruction manual. Each participant attended an experimental session of approximately 15 minutes and a post-interview session with a short self-report survey.
\subsection{Participants and Procedure}
We recruited 30 students with at least a bachelor's degree for the user study via the university bulletin board system.
The final sample consisted of 30 participants (25 males, 5 females), with age ranging from 16 to 45 years (average of 20.13).
About 27\% of them (8/30) had rich experience using brick products and AR devices, while the rest of them (22/30) had little.
In addition, 40 \% of participants (12/30) knew about natural language processing and machine learning. {Considering the direct comparison between our system and the traditional printed instruction is unfair, we considered a cross-group study with three conditions. The participants were randomly (lottery)} assigned to one of three test scenario groups: \textbf{ARAI}: AR device with AI generates sequence; \textbf{ARM}: AR device with manual adjusted sequence; \textbf{PM}: traditional printed instruction book with manual sequence. Each group consisted of 10 samples. 

The participants first attended a tutorial session, to assemble a simple model with 5 to 15 bricks in order to get familiar with our system. The tutorial session took around 5 minutes.
Then the main experiment started. 
The task is to assemble a brick set,
which consisted of 22 procedural steps. With the help of the aforementioned system, in every step, participants fetched a specific brick from the remaining bricks and put it in the correct position to complete the assembly tasks. {During the experiment, the user study moderator recorded the time of completion and the incorrect assembling trials in the experiment. After completing all tasks, the moderator asked the participants to fill out a survey and conduct one-on-one interviews about their experiences.}

    
{The primary goal of this study is to investigate whether the AR-based assemby guidance outperforms traditional assembly instructions. Additionally, we were also curious about the differences between AI generated assembly sequences and manual ones.} Therefore, we adopted framework including quantitative measurement and qualitative semi-structured interviews. The quantitative study included measurements on the effectiveness of different approaches and an adapted version of the User Experience Questionnaire Short version (UEQ-S)~\cite{schrepp2017design}. UEQ-S measures the User Experience of interactive products. Effectiveness is reflected by the time it takes to complete the task and by the amount of errors along the way. For example, selecting a component with the wrong color or shape, installing a piece at the wrong location or in the wrong orientation, and skipping a step. The qualitative study focuses on the pragmatic and hedonic dimensions, assessed through follow-up semi-structured interviews. We adapted the interview structure of~\citet{EggerLampl2019AssemblyIW}. The pragmatic dimension was measured by assistance, controllability, and willingness to try AR applications or brick models in the future. The hedonic dimension was evaluated by immersion, accomplishment, and enjoyment while playing.

\subsection{Apparatus And Software}

We programmed and ran the BrickPal system on a laptop computer with an 8-core AMD 5800H CPU and an NVIDIA GTX 1650 GPU.
The display hardware is a commercial AR HMD with binocular 2K resolution, 50 FOV, 6DOF spatial positioning, and 105 forward FOV. Both ARAI and ARM groups used the same device.
For the physical brick model in our user test, we use a commercial brick set designed for a cartoon avatar. The ARAI and ARM groups used the same hardware: AR HMD and laptop, while the PM group used printed paper instructions.
\subsection{Results and Discussions}
\begin{wrapfigure}{r}{0.25\textwidth}
\vspace{-2em}
    \centering
    \includegraphics[width=0.25\textwidth]{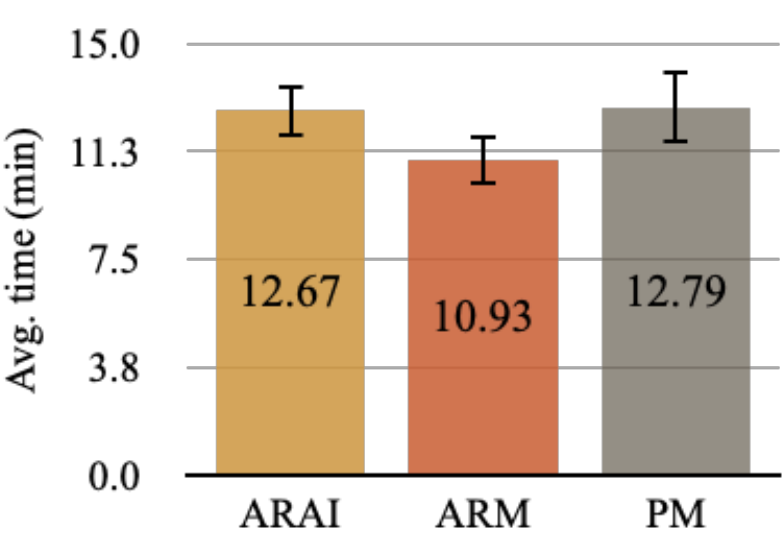}
   \\
    \includegraphics[width=0.25\textwidth]{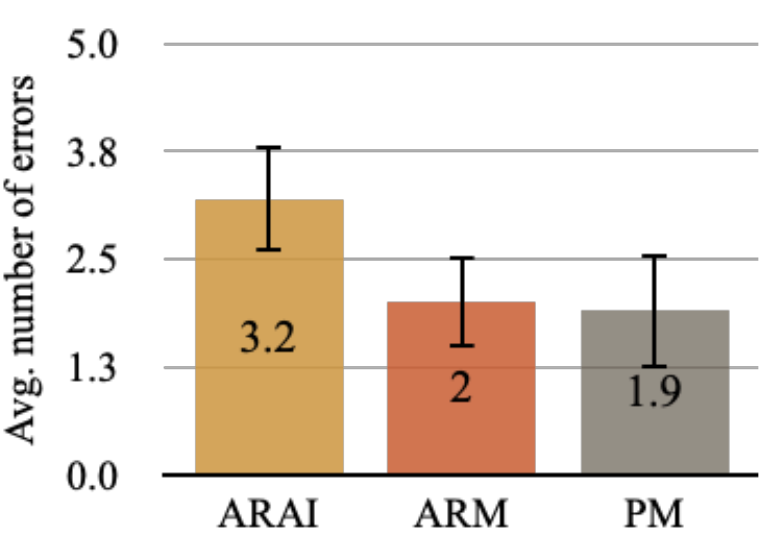}
\vspace{-3em}
\end{wrapfigure} 

\subsubsection{Efficiency And Effectiveness}

We conducted a one-way ANOVA to compare the efficiency of the three groups. The results ($p=0.371$) indicated similarity in the three treatments: ARAI, ARM, and PM. The right figure (up) revealed the average time of completion of each treatment. Participants in the ARM group spent the shortest time, while participants in the PM group took the longest time during the experiments. And the average consumed time in ARAI group was longer compared to ARM group.
The time among these three groups varies little. ARAI spent more time than ARM as ARAI is a step-by-step instruction while ARM can assemble several bricks in one step.

We also compared its effectiveness with adopting the same measure. The results ($p=0.256$) revealed that the differences between the groups were not statistically significant between the groups. The right figure (down) illustrated the average number of errors during the experiments (22 steps in total). The descriptive statistics revealed that the printed manual group has the lowest error rate in all categories. By contrast, the subjects in the ARAI group made the most mistakes. 
This result was due to the limited conditions of the VR equipment we applied, resulting in a higher overall error rate compared to paper instructions. 


\subsubsection{UEQ}
\begin{figure}[!tbp]
    \centering
    \includegraphics[width=0.8\linewidth]{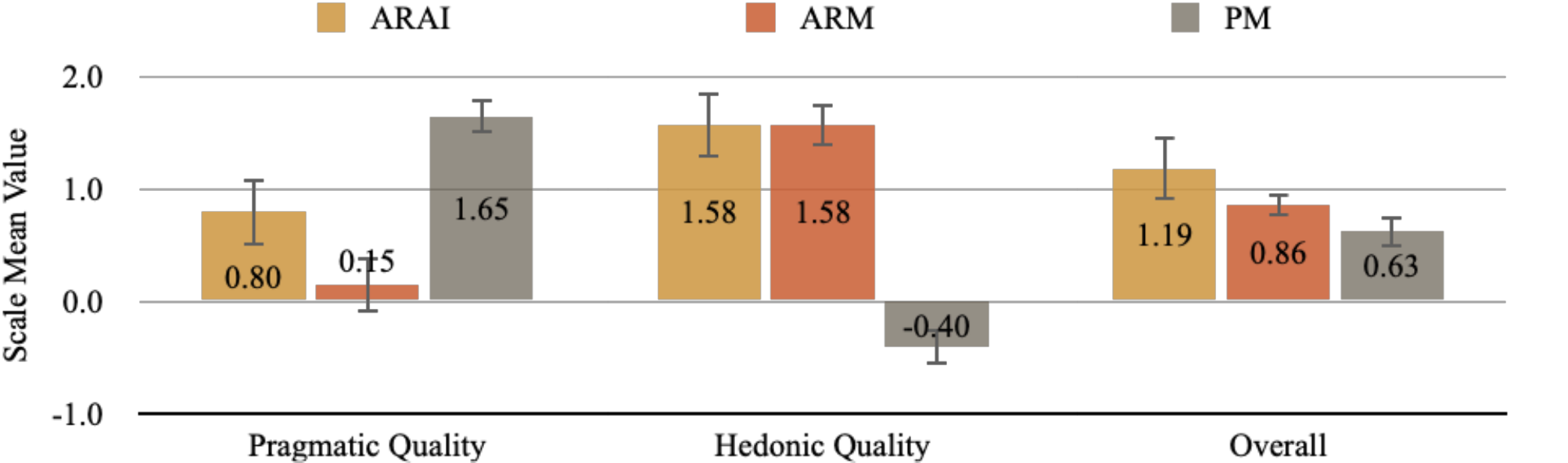}
    \caption{The UEQ-S results in 3 groups(ARAI, ARM, PM) in the quality categories of pragmatic, hedonic and overall scores. Values > 0,8 represent a positive evaluation and values < -0,8 represent a negative evaluation.}
    \label{fig:ueq}
    \vspace{-1.2em}
\end{figure}

As introduced by \citet{schrepp2017design}, the UEQ-S (User Experience Questionnaire-short) scale is a widely used scale for analyzing user experience. The scale range is between -3 (horribly bad) and +3 (extremely good). We adapted this scale to measure the user experience of our system. As shown in Fig. \ref{fig:ueq}, it revealed that ARAI treatment has the highest overall score. While the other two groups gave a below-average score compared to the benchmark(50\% results are better and 25\% results are worse). The scale of hedonic quality is positive in each of groups with AR, i.e. the ARAI group and the ARM group. However, the PM group presents a better score with pragmatic quality. {The scale of pragmatic quality involved descriptive words such as easy and efficient.
The possible reasons include: Most of the participants had little experience in AR, and 11 (36.6\%) of them had never assembled bricks before. Thus, the AR instructions were not easy to follow.}


\subsubsection{Semi-structured Interview}
We used an in-depth qualitative analysis method~\cite{Glaser1968TheDO} to code and interpret the data collected from interviews. We concluded the interview with four key themes regarding participants' perceptions, and discuss the functional and emotional aspects of the system:

\textit{\textbf{The task is more intuitive in AR.}} In terms of the helpfulness of the system, compared to traditional printed paper manuals, our AR system provides a more direct and clearer task guidance. 17 participants (85\%) reported this. The presentation primarily reflects intuitive step guidance, accurate position indication, and clear component features. For example, P7 (AR+AI group) explained the most helpful and supportive part of his experience: \textit{``This tool led me straight to the next step, specifically moving the base up to a 3D model, as I found it very clear from multiple angles of view.''}  P6 (AR+M group) also said,\textit{``The guidance with a 3D model, which introduced me to the brick I needed to pick up and where the place was. I could even turn it around to observe from a different angle of view. By the way, the 3D model has high precision.''}  However, there is also a voice from the other side. 6 participants indicated there is space for improvement for our virtual model. For instance, the color chromatic aberration btw virtual/physical parts confused them, and the green-line model was unclear after steps accumulated.

\textit{\textbf{Mixed Reality makes the assembling process playful.}} We asked participants the question, \textit{``Do you feel pleased when interacting with the system? such as virtual bubble interaction.''}  We received 13 (65\%) positive answers. They had fun building bricks with AR guidance and interacting with our system. For instance, P9 told us,\textit{``It gave me a lot of freshness, touching the bubble to unlock the next step and use a magnifier to get details. I believe it was much more interesting than just following a printed document.''} He was not the only one with this opinion. P10 also said, \textit{"In my opinion, the sound effect of pricking bubbles was so funny, and the magnifier was not bad as well. I like the feeling of doing this."}  Participants shared the familiar feeling of pleasure and fun through assembling with BrickPal, and the reasons revealed that the fun was issued from different aspects, such as learning how to use an AR device, interacting with virtual objects. By comparison, 7 participants indicated they had no fun with traditional instructions. They explained that the printed instruction itself is not interesting, but searching for components is pleasant.


\textbf{\textit{Comparison of AI-generated assembly sequences and manual adjusted sequences.}}
Based on the interview, we found that participants were unlikely to recognize the difference between AI-generated sequences and manual production. The answers about sense of achievement and sense of difficulty were similar in both groups. However, after they learned about the flexible sequencing scheme which allows multiple tracks, they agreed to a potentially higher degree of freedom and greater willingness to explore bricks assembling with AI.

\section{Conclusion}
This paper presented BrickPal, an AR system for assisting users assemble bricks.
With machine learning-based assembly sequence generation,
our system achieved interactive real-time generation of assembly sequences for brick models.
The numerical evaluation and user study on comparison of our system with traditional assembly instructions shown the robustness of our system and the enhanced experience.
In the future, we will focus on improving the 3D-PALM algorithm and exploring different user interactions between the AR system and the 3D-PALM model.
\clearpage


\bibliographystyle{ACM-Reference-Format}
\bibliography{references}


\end{document}